\documentclass[conference]{IEEEtran}
\IEEEoverridecommandlockouts

\usepackage{cite}
\usepackage{amsmath,amssymb,amsfonts}
\usepackage{algorithmic}
\usepackage{graphicx}
\usepackage{textcomp}
\usepackage{fontawesome5} 
\usepackage{flushend}
\usepackage{booktabs}
\usepackage{xspace}
\newcommand{\partitle}[1]{\noindent{\textbf{#1.}}}
\newcommand{\ours}{DiTSG\xspace}

\usepackage[square, numbers, sort]{natbib}
\usepackage[table, dvipsnames]{xcolor}
\usepackage[font=small, labelfont=bf]{caption}
\usepackage[colorlinks = true,
            linkcolor = blue,
            urlcolor  = blue,
            citecolor = magenta,
            bookmarks = true, pagebackref]{hyperref}

\def\BibTeX{{\rm B\kern-.05em{\sc i\kern-.025em b}\kern-.08em
    T\kern-.1667em\lower.7ex\hbox{E}\kern-.125emX}}
\begin{document}

\title{\LARGE \bf \ours: Context-Aware Air Traffic Scene Generation \\ with Multi-Agent Diffusion Transformers}

\author{\IEEEauthorblockN{Seokbin Yoon$^{1}$, Kyungmin Kim$^2$, Taejung Kwon$^2$, Keumjin Lee$^2$, Peng Wei$^{1}$}
\IEEEauthorblockA{$^1$George Washington University, Washington, DC 20052, USA}
\IEEEauthorblockA{$^2$Korea Aerospace University, Goyang 10540, South Korea}
\{seokbin.yoon, pwei\}@gwu.edu, \{kyungmin.kim, ktj0302\}@kau.kr, keumjin.lee@kau.ac.kr}

\maketitle
\begin{abstract}
Understanding how many aircraft, and which traffic configurations, can be safely and efficiently managed by air traffic controllers within a given airspace is fundamental to air traffic management. Air traffic controllers continuously modify aircraft trajectories through interventions such as speed advisories and radar vectoring, and the patterns of these interventions are difficult to capture effectively using existing air traffic simulators. To address this limitation, we propose Diffusion Transformers for Traffic Scene Generation (DiTSG), which learns to generate air traffic scenes from recorded multi-aircraft trajectories. DiTSG employs stacked multi-agent inverted Transformer layers as the denoising backbone of a diffusion transformer, jointly generates the trajectories of multiple aircraft within a traffic scene, and conditions the generation on agent-specific required times of arrival. The resulting trajectories generated by DiTSG preserve realistic inter-aircraft interactions observed in actual air traffic operations. Experiments show that DiTSG generates diverse traffic scenes that reproduce the statistical properties of recorded traffic while preserving realistic inter-aircraft interactions. The project page and source code are available at
\href{https://seokbinyoon96.github.io/ditsg/}{\textcolor{magenta}{https://seokbinyoon96.github.io/ditsg/}}.
\end{abstract}


\section{Introduction}
Modeling and simulating interacting flight trajectories in airspace is fundamental to assessing the safety and efficiency of aircraft operations, including the maximum level of traffic complexity that air traffic management (ATM) systems can safely accommodate~\cite{schmidt1976modeling,majumdar2001estimating,welch2015route}. A key challenge is that air traffic controllers (ATCs) continuously modify aircraft trajectories through interventions such as radar vectoring and speed changes, both to maintain safe separation and to improve flight time and fuel consumption~\cite{saez2020traffic,hong2015trajectory}. Because these interventions depend on various factors, including the surrounding traffic, the real-world aircraft trajectories can unfold in many different ways even under similar levels of traffic demand. Consequently, no single traffic scenario can adequately represent the range of possible traffic situations. An air traffic generation model should therefore capture the distribution of possible realizations rather than generate only a single set of trajectories for a given scenario.

A common approach to generating traffic is to use air traffic simulators such as BlueSky~\cite{hoekstra2016bluesky}. Although this approach enables a large-scale simulation of air traffic operations based on predefined flight plans and airspace procedures, realistically reproducing how ATCs modify aircraft trajectories remains challenging. This limitation is particularly important in terminal airspace, where controller interventions are frequent and substantial. As a result, conventional simulation-based approaches may have limited ability to reproduce complex traffic behaviors that emerge from real-world ATC operations.

Alternatively, data-driven generative modeling can learn air traffic behavior directly from real-world trajectories. From a probabilistic perspective, each observed traffic scene can be regarded as a sample from an underlying, generally unknown distribution of possible air traffic operations. A generative model can learn this distribution and then generate new, plausible samples beyond those in the observed dataset, broadening the range of simulated traffic scenarios.

Existing data-driven trajectory generation methods primarily focus on modeling individual aircraft behavior~\cite{yoon2026aircraft,krauth2023deep,yoon2023improving,pepper2025sector}. However, a trajectory generated in isolation cannot preserve how ATCs perceive and manage the traffic situation as a whole. Consequently, independently sampling individual trajectories may yield a traffic scene with a collection of individually plausible flights, but does not necessarily produce a realistic traffic scene that preserves relationships among aircraft, such as separation and sequencing. Generating traffic scenes in which multiple aircraft properly interact is therefore required to capture not only individual flight behavior but also inter-aircraft dependencies. 

Toward this goal, we propose Diffusion Transformers for Traffic Scene Generation (\textit{\ours}), a context-aware generative model for multi-agent air traffic scenes, where all aircraft trajectories are generated jointly and each trajectory is modeled in the context of the others. \ours employs stacked multi-agent inverted Transformer (MAIFormer)~\cite{yoon2026multi} layers as the denoising backbone of a diffusion transformer. Generation is further conditioned on agent-specific required times of arrival (RTAs), enabling different temporal traffic configurations. Starting from noise, \ours iteratively refines the trajectories of all aircraft in a scene while modeling inter-aircraft dependencies at every denoising step. This joint denoising process enables the generated trajectories to remain mutually consistent while preserving realistic traffic interactions.

Our main contributions are threefold:
\begin{enumerate}
    \item We propose \ours, a context-aware generative framework that formulates air traffic scene generation as a joint multi-agent generation problem, where each aircraft trajectory is generated in the context of the others rather than being modeled independently from the surrounding traffic.

    \item We develop a diffusion transformer with a multi-agent denoising backbone that enables information aggregation across aircraft through attention at every denoising step, while supporting a variable number of aircraft and controllability through agent-specific RTAs.
    
    \item We evaluate \ours using real-world terminal airspace data and demonstrate that it generates diverse traffic scenes under the same conditional information and maintains realistic inter-aircraft separation.
\end{enumerate}

\section{Related Work}\label{related}
This work draws on two lines of research: aircraft trajectory generation and multi-agent learning for ATM systems. We review each area below and discuss their limitations for context-aware multi-agent air traffic scene generation.

\subsection{\textbf{Trajectory Generation}}
Before the emergence of data-driven approaches, aircraft trajectory generation largely relied on physics-based models, in which aircraft motion is governed by aerodynamic principles and performance models~\cite{nuic2010bada,sun2020openap}. These models provide strong physical fidelity, but have limited ability to reproduce the uncertainty and variability observed in real-world trajectories. Consequently, they are less suited to generating diverse traffic realizations under different operational conditions~\cite{olive2021framework}. One of the efforts toward overcoming this limitation therefore lies in data-driven trajectory generation, in which models learn uncertainty and variability directly from the archived real-world trajectory data.

Early data-driven studies used statistical methods such as Gaussian mixture models (GMMs)~\cite{barratt2018learning,jung2025inferring} and synthetic minority oversampling~\cite{yoon2023improving} for trajectory generation. Deep generative approaches later combined temporal convolutional networks (TCNs) with variational autoencoders (VAEs) for trajectory generation~\cite{krauth2023deep}, including joint generation of trajectory pairs for collision risk assessment~\cite{krauth2024advanced}. Generative adversarial networks (GANs) have also been applied to synthesize flight trajectories~\cite{jarry2019use,wijnands2024generation}, while transformer-based autoencoders have been paired with GMMs in the latent space~\cite{yoon2026aircraft}. More recently, diffusion and flow matching models have been explored for trajectory generation~\cite{matsuda2026high}.

Prior work, however, primarily focuses on individual aircraft or limited pairwise interactions, which cannot fully capture the joint behavior of a traffic scene. \ours addresses this limitation by jointly generating multiple aircraft while explicitly modeling inter-aircraft dependencies. It also supports generating a variable number of aircraft, reflecting the fact that the number of aircraft simultaneously present within an airspace naturally varies over time.

\subsection{\textbf{Multi-Agent Learning}}
ATM systems are inherently multi-agent, since ATC interventions affect the trajectories of several interacting aircraft, but multi-agent learning remains less explored than its single-agent counterpart in the ATM domain. Social long short-term memory (LSTM) networks~\cite{xu2021multi} and attention-based models~\cite{nguyen2025multi} capture inter-aircraft interactions for trajectory prediction, while transformers combined with Bayesian neural networks jointly model interactions and uncertainty in traffic scenes~\cite{pang2022bayesian}. Agent-aware attention inspired by AgentFormer~\cite{yuan2021agentformer} has also been adapted to air traffic scenes~\cite{deng2025multi}. MAIFormer was proposed more recently, using a hierarchical attention mechanism that mimics ATCs' decision-making for context-aware modeling of interacting aircraft~\cite{yoon2026multi}.

Nevertheless, prior multi-agent learning approaches primarily focus on supervised prediction of future trajectories or flight times and are therefore not designed to generate novel traffic scenarios beyond the observed samples, limiting their ability to explore diverse plausible traffic configurations. \ours extends multi-agent learning from prediction to generation by jointly modeling interacting aircraft and producing diverse traffic realizations that remain consistent with patterns learned from observed data.

\section{Preliminaries}\label{prelim}
\subsection{\textbf{Diffusion and Flow Matching}}
Diffusion-based generative models transport samples from a simple prior distribution to the data distribution through a continuous-time process described by differential equations~\cite{ho2020denoising}. In this work, we adopt the ordinary differential equation (ODE) formulation and train our diffusion transformer with a conditional flow matching (CFM) objective on rectified linear interpolation paths~\cite{lipman2023flow,liu2023flow}. CFM directly learns a velocity field that transports samples along a prescribed probability path.

Let $\mathbf{x}_0 \sim p_0$ denote a sample from the prior distribution, where 
$p_0=\mathcal{N}(\mathbf{0},\mathbf{I})$, and let $\mathbf{x}_1 \sim p_1$ denote a 
sample from the data distribution. A continuous probability path connects $p_0$ and 
$p_1$ over the diffusion timestep $s\in[0,1]$. We define 
this path using linear interpolation:
\begin{equation}
\mathbf{x}_s
=
(1-s)\mathbf{x}_0
+
s\mathbf{x}_1 .
\end{equation}
The path transports each paired noise and data sample along a straight line from $\mathbf{x}_0$ at $s=0$ to $\mathbf{x}_1$ at $s=1$, so the target velocity is constant and given by
\begin{equation}
\mathbf{v}_s
=
\frac{d\mathbf{x}_s}{ds}
=
\mathbf{x}_1-\mathbf{x}_0 .
\end{equation}
CFM learns a parameterized velocity field $\mathbf{v}_\theta(\mathbf{x}_s,s)$ that 
approximates this target velocity by minimizing
\begin{equation}
\mathcal{L}_{\mathrm{CFM}}(\theta)
=
\mathbb{E}_{\substack{
s\sim\mathcal{U}(0,1),\\
\mathbf{x}_0\sim p_0,\;
\mathbf{x}_1\sim p_1
}}
\left[
\left\|
\mathbf{v}_\theta(\mathbf{x}_s,s)
-
(\mathbf{x}_1-\mathbf{x}_0)
\right\|_2^2
\right].
\end{equation}
Once trained, the learned velocity field defines the ODE
\begin{equation}
\frac{d\mathbf{x}_s}{ds}
=
\mathbf{v}_\theta(\mathbf{x}_s,s).
\end{equation}
We generate a sample by integrating this ODE from 
$\mathbf{x}_0\sim\mathcal{N}(\mathbf{0},\mathbf{I})$ at $s=0$ toward $s=1$. Using a 
first-order Euler solver, each integration step is
\begin{equation}
\mathbf{x}_{s+\Delta s}
=
\mathbf{x}_s
+
\Delta s \,
\mathbf{v}_\theta(\mathbf{x}_s,s).
\end{equation}

\subsection{\textbf{Multi-Agent Inverted Transformer}}\label{maiformer}
MAIFormer is a multi-agent model for air traffic scenes that captures intra-agent trajectory evolution at the low level and inter-agent influence at the high level~\cite{yoon2026multi}. It has proven a versatile backbone for multi-agent modeling, with subsequent extensions to arrival time prediction~\cite{kim2026probabilistic} and to interpreting air traffic situations through agent attention scores~\cite{chai2025learning}. MAIFormer is an encoder-only architecture with three key components: (i) air traffic scene tokenization, (ii) masked multivariate attention (MMA), and (iii) agent attention (AA)\footnote{We refer readers to~\cite{yoon2026multi} for full architectural details.}.

\begin{figure*}[t!]
    \centering
    \includegraphics[width=1.0\textwidth]{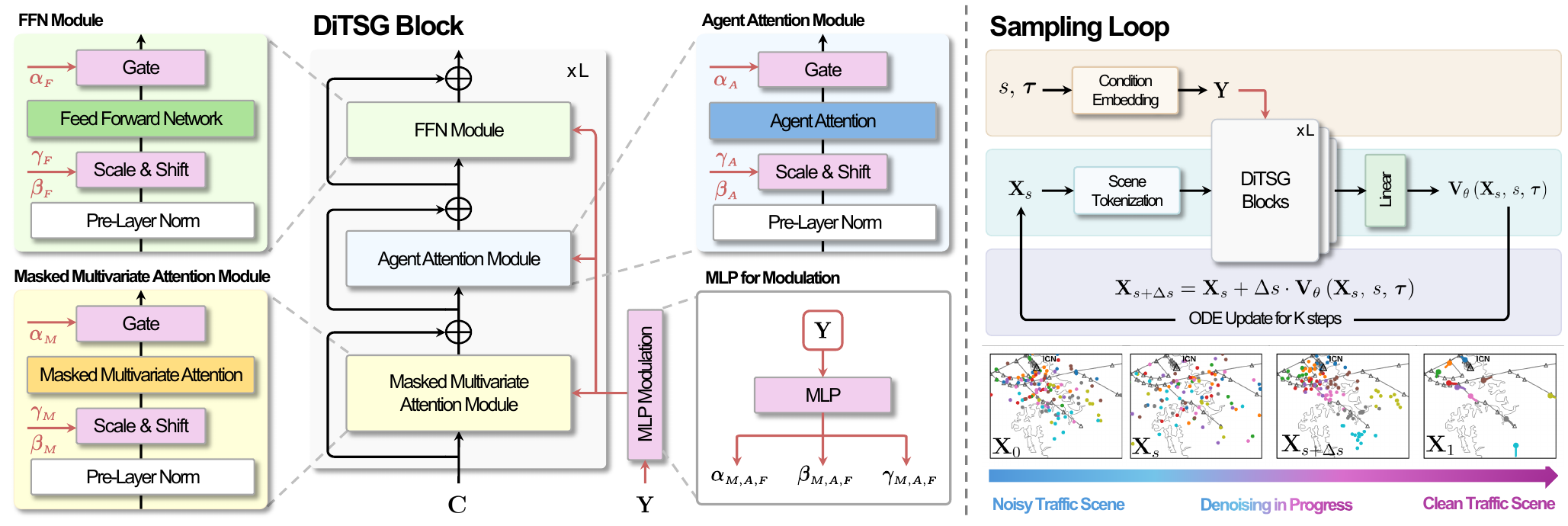}
    \vspace{-15pt}
    \caption{\textbf{Overview of \ours}. Left: a \ours block, which applies masked multivariate attention, agent attention, and a feed-forward layer, each modulated by the conditioning vector $\mathbf{Y}$ through adaLN. Right: the sampling loop, which tokenizes the noisy scene $\mathbf{X}_s$, passes it through $L$ blocks and a linear layer to obtain the velocity $\mathbf{V}_{\theta}$, and integrates the ODE for $K$ Euler steps from noise to a traffic scene.}
    \label{archi}
\end{figure*}

\vspace{3pt}
\partitle{Air Traffic Scene Tokenization}
MAIFormer's tokenization extends the inverted embedding strategy of iTransformer~\cite{liu2024itransformer}, which embeds along the variate dimension to form variate tokens rather than temporal tokens and has been effective for multivariate time-series and trajectory modeling~\cite{yoon2025aircraft}. Given a multi-agent flight trajectory sequence $\mathbf{X} \in \mathbb{R}^{N\times T\times F}$ ($N$ agents, $T$ time steps, $F$ variates), we reshape it into $\mathbf{\overline{X}} \in \mathbb{R}^{(N\cdot F) \times T}$ and treat the entire scene as a single multivariate time series. Scene tokenization then projects the $T$-dimensional temporal sequence of each variate into the model latent space of dimension $D$, producing
\begin{equation}
\mathbf{C} = \operatorname{SceneTokenization}(\mathbf{\overline{X}}) \in \mathbb{R}^{(N\cdot F) \times D},
\end{equation}
where $\mathbf{C}=(\mathbf{c}^1_1, \dots, \mathbf{c}^F_1, \dots, \mathbf{c}^1_N, \dots, \mathbf{c}^{F}_{N})$, and each token represents the temporal evolution of one variate for one aircraft.


\vspace{3pt}
\partitle{Masked Multivariate Attention}
MMA applies self-attention (SA)~\cite{vaswani2017attention} to the variate tokens $\mathbf{C}$ to model spatio-temporal (ST) dependencies within each aircraft, using a mask $\mathbf{M}$ that blocks cross-agent interactions, such that the tokens of each aircraft attend only to other variates of the same aircraft:
\begin{equation}
\begin{aligned}
&\operatorname{MMA}(\mathbf{Q}_{\mathrm{ST}},\mathbf{K}_{\mathrm{ST}},\mathbf{V}_{\mathrm{ST}}) = \operatorname{Softmax}\left(\frac{\mathbf{Q}_{\mathrm{ST}}\mathbf{K}_{\mathrm{ST}}^{\top}}{\sqrt{d_{k,\mathrm{ST}}}} + \mathbf{M}\right)\mathbf{V}_{\mathrm{ST}}, \\
&\mathbf{Q}_{\mathrm{ST}} = \mathbf{C}\mathbf{W}^{Q}_{\mathrm{ST}}, \quad \mathbf{K}_{\mathrm{ST}} = \mathbf{C}\mathbf{W}^{K}_{\mathrm{ST}}, \quad \mathbf{V}_{\mathrm{ST}} = \mathbf{C}\mathbf{W}^{V}_{\mathrm{ST}},
\end{aligned}
\end{equation}
where each $\mathbf{W}_{\mathrm{ST}} \in \mathbb{R}^{D \times D}$ projects $\mathbf{C}$ into queries, keys, and values. By restricting information flow to the variates of each individual aircraft, MMA captures intra-agent dependencies.

\vspace{3pt}
\partitle{Agent Attention}
AA models the social context (SC) among agents by applying SA across agent-level tokens. Each agent token is formed by concatenating the variate tokens produced by the preceding MMA stage:
\begin{equation}
\mathbf{C}_{\mathrm{A},i} = (\mathbf{c}_{i}^{1} \oplus \cdots \oplus \mathbf{c}_{i}^{F}) \in \mathbb{R}^{1 \times (F \cdot D)},
\end{equation}
where $\oplus$ denotes concatenation along the variate dimension. SA is then applied to $\mathbf{C}_{\mathrm{A}} \in \mathbb{R}^{N \times (F \cdot D)}$ with separate projections $\mathbf{W}_{\mathrm{SC}} \in \mathbb{R}^{(F\cdot D) \times (F\cdot D)}$:
\begin{equation}
\begin{aligned}
&\operatorname{AA}(\mathbf{Q}_{\mathrm{SC}},\mathbf{K}_{\mathrm{SC}},\mathbf{V}_{\mathrm{SC}}) = \operatorname{Softmax}\left(\frac{\mathbf{Q}_{\mathrm{SC}}\mathbf{K}_{\mathrm{SC}}^{\top}}{\sqrt{d_{k,\mathrm{SC}}}}\right)\mathbf{V}_{\mathrm{SC}}, \\
&\mathbf{Q}_{\mathrm{SC}} = \mathbf{C}_{\mathrm{A}}\mathbf{W}^{Q}_{\mathrm{SC}}, \quad \mathbf{K}_{\mathrm{SC}} = \mathbf{C}_{\mathrm{A}}\mathbf{W}^{K}_{\mathrm{SC}}, \quad \mathbf{V}_{\mathrm{SC}} = \mathbf{C}_{\mathrm{A}}\mathbf{W}^{V}_{\mathrm{SC}}.
\end{aligned}
\end{equation}
AA therefore relates whole aircraft rather than individual variates, at a level of abstraction consistent with how human ATCs reason about air traffic scenes.

\section{Technical Approach}\label{approach}
\ours generates multi-agent air traffic scenes by learning a joint velocity field over aircraft trajectories. We use a stack of $L$ MAIFormer layers to parameterize this field, allowing inter-aircraft dependencies to be modeled throughout the generation process. Traffic scene generation is conditioned on the RTAs $\boldsymbol{\tau}$ of all aircraft. Figure~\ref{archi} provides an overview of \ours.

\subsection{\textbf{\ours Neural Architecture}}
\partitle{Noisy Scene Tokenization}
Let $\mathbf{X}_1 \in \mathbb{R}^{N\times T\times F}$ denote a real traffic scene. During training, we sample a diffusion timestep $s\sim\mathcal{U}(0,1)$ and noise $\mathbf{X}_0 \in \mathbb{R}^{N\times T\times F}\sim\mathcal{N}(\mathbf{0},\mathbf{I})$, and construct the interpolated scene $\mathbf{X}_s=(1-s)\mathbf{X}_0+s\mathbf{X}_1$. We reshape $\mathbf{X}_s$ into $\overline{\mathbf{X}}_s$ and apply scene tokenization:
\begin{equation}
\mathbf{C} = \operatorname{SceneTokenization}(\overline{\mathbf{X}}_s) \in \mathbb{R}^{(N\cdot F)\times D}.
\end{equation}
The tokens are those of a partially noised scene, so what a token encodes changes with $s$: near $s=1$ it carries the variate's actual time series, and near $s=0$ it carries little beyond noise.

\partitle{Agent-wise Conditioning}
\ours conditions the generation on two quantities. The diffusion timestep $s$ applies to the entire scene and indicates its progression from noise toward data. The RTA $\tau_i \in \mathbb{R}$ is an agent-specific control input that specifies the time remaining until aircraft $i$ is required to arrive at the runway. During generation, the RTAs are specified externally and may be provided by an arrival scheduler. Across agents, $\boldsymbol{\tau} \in \mathbb{R}^{N}$ specifies the relative arrival order and temporal spacing among aircraft. Given this RTA configuration, \ours generates plausible traffic scenes that could occur under the specified arrival order and spacing. We embed both quantities using sinusoidal encodings followed by multi-layer perceptrons (MLPs) and combine them into one conditioning vector for each agent:
\begin{equation}
\mathbf{y}_i = \operatorname{MLP}^{s}\!\left(\operatorname{SinEnc}(s)\right) + \operatorname{MLP}^{\tau}\!\left(\operatorname{SinEnc}(\tau_i)\right) \in \mathbb{R}^{D},
\label{eq:cond}
\end{equation}
collected across agents as $\mathbf{Y} = (\mathbf{y}_1,\dots,\mathbf{y}_N) \in \mathbb{R}^{N\times D}$. Following DiT~\cite{peebles2023scalable}, we inject $\mathbf{Y}$ through adaptive layer normalization, $\operatorname{Mod}(\mathbf{c},\boldsymbol{\gamma},\boldsymbol{\beta}) = (1+\boldsymbol{\gamma})\odot\operatorname{LN}(\mathbf{c})+\boldsymbol{\beta}$, at every sublayer. An MLP maps each $\mathbf{y}_i$ to a shift, scale, and gate:
\begin{equation}
\{\boldsymbol{\beta}_i^{k},\boldsymbol{\gamma}_i^{k},\boldsymbol{\alpha}_i^{k}\}_{k\in\{M,A,F\}} = \operatorname{MLP}^{\mathrm{mod}}(\mathbf{y}_i),
\end{equation}
where $k$ indexes the MMA, AA, and FFN sublayers. The shift and scale are applied through $\operatorname{Mod}$, while the gate scales each sublayer output before it is added back to the residual stream. These modulation parameters are shared across all tokens of aircraft $i$, providing agent-specific conditioning throughout the network.

\vspace{3pt}
\partitle{Denoising Backbone}
Each \ours block applies the three MAIFormer sublayers with adaLN replacing the original layer normalization:
\begin{equation}
\begin{aligned}
\mathbf{C} &= \mathbf{C} + \boldsymbol{\alpha}^{M}\odot\operatorname{MMA}(\operatorname{Mod}(\mathbf{C},\boldsymbol{\gamma}^{M},\boldsymbol{\beta}^{M})), \\
\mathbf{C}_{\mathrm{A}} &= \mathbf{C}_{\mathrm{A}} + \boldsymbol{\alpha}^{A}\odot\operatorname{AA}(\operatorname{Mod}(\mathbf{C}_{\mathrm{A}},\boldsymbol{\gamma}^{A},\boldsymbol{\beta}^{A})), \\
\mathbf{C} &= \mathbf{C} + \boldsymbol{\alpha}^{F}\odot\operatorname{FFN}(\operatorname{Mod}(\mathbf{C},\boldsymbol{\gamma}^{F},\boldsymbol{\beta}^{F})).
\end{aligned}
\end{equation}
MMA attends within each agent's own tokens and AA attends across agents, while the feedforward neural network (FFN) applies a nonlinear transformation to each token. Because AA is applied in every block, each aircraft is generated in the context of the surrounding traffic throughout the denoising process, rather than independently.

\vspace{3pt}
\partitle{Velocity Prediction}
A final adaLN modulates the output tokens $\mathbf{C}_{\mathrm{out}} \in \mathbb{R}^{(N\cdot F)\times D}$, after which a linear layer projects each token back to its $T$-step variate sequence:
\begin{equation}
\overline{\mathbf{V}}_{\theta} = \operatorname{Linear}\!\left(\operatorname{Mod}(\mathbf{C}_{\mathrm{out}},\boldsymbol{\gamma}^{\mathrm{out}},\boldsymbol{\beta}^{\mathrm{out}})\right) \in \mathbb{R}^{(N\cdot F)\times T}.
\end{equation}
Reversing the reshape used during tokenization yields the predicted velocity field $\mathbf{V}_{\theta}(\mathbf{X}_s,s,\boldsymbol{\tau}) \in \mathbb{R}^{N\times T\times F}$ over the multi-agent traffic scene.

\subsection{\textbf{Training Objective and Traffic Scene Generation}}
We train \ours using the CFM objective, conditioned on RTAs $\boldsymbol{\tau}$:
\begin{equation}
\mathcal{L}_{\mathrm{CFM}}(\theta)
=
\mathbb{E}_{\substack{s\sim\mathcal{U}(0,1),\\ \mathbf{X}_1\sim p_1,\; \mathbf{X}_0\sim p_0}}
\left[
\left\|
\mathbf{V}_{\theta}(\mathbf{X}_s,s,\boldsymbol{\tau})-(\mathbf{X}_1-\mathbf{X}_0)
\right\|_2^2
\right].
\end{equation}
At inference, we specify the number of aircraft $N$ and their RTAs $\boldsymbol{\tau} \in \mathbb{R}^{N}$, sample $\mathbf{X}_0 \in \mathbb{R}^{N\times T\times F} \sim\mathcal{N}(\mathbf{0},\mathbf{I})$, and integrate the learned ODE from $s=0$ to $s=1$ using a $K$-step Euler solver. Because \ours does not require a fixed $N$, the same model can generate traffic scenes with different numbers of aircraft, reflecting the varying number of aircraft that may be present within an airspace.

\section{Experimental Setup}\label{setup}
\subsection{\textbf{ADS-B Trajectory Dataset Description}}
We evaluate \ours on real-world ADS-B trajectories from the terminal airspace around Incheon International Airport (ICN), South Korea, covering arrivals to Runway 15L from January to December 2024. As one of the world's busiest airports, ICN makes extensive use of area navigation (RNAV), allowing flexible flight-path adjustments under varying traffic conditions. This produces diverse inter-aircraft interactions, making ICN a suitable testbed for traffic scene generation.

We construct scenes using a two-minute sliding window advancing by one 6-second time step, yielding $T=20$ steps per scene. Each scene contains the trajectories of all aircraft present during the corresponding interval. We focus on short-horizon scene generation and leave longer horizons to future work. This yields 991,665 scenes, which we split chronologically into training and test sets with a 9:1 ratio.

\subsection{\textbf{Model and Training Configurations}}
\ours stacks $L=6$ MAIFormer layers with a hidden dimension of $D=128$ and an FFN dimension of 512 ($4D$), using GELU activations~\cite{hendrycks2016gaussian}. Each aircraft trajectory consists of $F=3$ variates: latitude, longitude, and altitude. We use four attention heads for both MMA and AA.

We train \ours for 300 epochs with a batch size of $1024$ using AdamW~\cite{loshchilov2019decoupled} and a learning rate of $10^{-4}$. At inference, we integrate the ODE using a 50-step Euler solver. All experiments are implemented in Python 3.11.16 with PyTorch 2.5.1~\cite{paszke2019pytorch} and CUDA 12.1. Experiments are run on Ubuntu 24.04 LTS using an Intel Xeon W-3245 CPU, 62 GB of RAM, and an NVIDIA RTX A6000 GPU.

\subsection{\textbf{Evaluation Metrics}}
We evaluate the generated traffic scenes along four dimensions: fidelity to real traffic, novelty, diversity, and safety compliance.

\vspace{3pt}
\partitle{Spatial Distribution Fidelity}
We evaluate whether \ours generates traffic in the same regions of airspace as real aircraft. We discretize the terminal airspace into a latitude--longitude grid and compute the Jensen--Shannon divergence (JSD) between the occupancy distributions $P_{\mathrm{real}}$ and $P_{\mathrm{gen}}$ of real and generated positions:
\begin{equation}
D_{\mathrm{occ}} = \operatorname{JSD}\left(P_{\mathrm{real}} \,\Vert\, P_{\mathrm{gen}}\right).
\label{eq:jsd}
\end{equation}
A lower $D_{\mathrm{occ}}$ indicates greater similarity between the spatial distributions of real and generated traffic.

\vspace{3pt}
\partitle{Novelty}
A model could achieve high fidelity by reproducing training scenes, so we also measure how far generated scenes lie from the training set. For two scenes with the same $N$, we match their aircraft one to one and take the assignment with the minimum average trajectory distance:
\begin{equation}
d(A, B) = \min_{\pi \in \Pi_N} \frac{1}{N}\sum_{i=1}^{N} \left\| \mathbf{x}_i^{A} - \mathbf{x}_{\pi(i)}^{B} \right\|,
\label{eq:hungarian}
\end{equation}
where $\mathbf{x}_{i}^A$ is the trajectory of aircraft $i$ in scene $A$ and $\Pi_N$ is the set of all permutations of $N$ aircraft. Defining $d(\cdot, \mathrm{Train}) = \min_{s \in \mathrm{Train}} d(\cdot, s)$ as the distance to the nearest training scene, we report
\begin{equation}
R_{\mathrm{nov}} =
\frac{\frac{1}{|\hat{S}|}\sum_{\hat{s}\in \hat{S}} d(\hat{s}, \mathrm{Train})}
{\frac{1}{|S^{\mathrm{test}}|}\sum_{s^{\mathrm{test}}\in S^{\mathrm{test}}} d(s^{\mathrm{test}}, \mathrm{Train})},
\label{eq}
\end{equation}
where $\hat{S}$ and $S^{\mathrm{test}}$ denote the generated and held-out test scenes, respectively. Because the test scenes represent unseen real traffic, they provide a reference for novelty: $R_{\mathrm{nov}}$ near one indicates that generated scenes are approximately as distant from the training data as unseen real scenes.

\vspace{3pt}
\partitle{Diversity}
We measure variation among traffic scenes generated under the same condition. For a fixed $\boldsymbol{\tau}$, we generate $K$ scenes and compute the average pairwise horizontal distance among them in nautical miles. We interpret this metric together with $D_{\mathrm{occ}}$, since greater diversity is meaningful only when generated traffic remains consistent with the spatial distribution of real traffic.

\vspace{3pt}
\partitle{Separation Violation Rate}
We evaluate safety compliance using the separation violation rate (SVR), defined as the fraction of scenes containing at least one pair of aircraft that simultaneously falls below 3 NM of horizontal separation and 1,000 ft of vertical separation. We report SVR for both generated and real test scenes to assess whether \ours preserves separation behavior observed in real traffic.

\section{Numerical Results}\label{results}

\subsection{\textbf{Quantitative Results}}
Table~\ref{tab:eval} compares \ours with a multi-agent VAE (MA-VAE) using the same MAIFormer backbone, so the main difference lies in the generative formulation: flow matching for \ours and variational latent modeling for MA-VAE.
\begin{table}[h]
\centering
\caption{Evaluation of generated traffic scenes.}
\vspace{-3pt}
\label{tab:eval}
\resizebox{0.8\columnwidth}{!}{%
\begin{tabular}{lcccc}
\toprule
Method & $D_{\mathrm{occ}}$ & $R_{\mathrm{nov}}$ & Diversity & SVR (\%) \\
\midrule
Real & -- & 1.000 & -- & 0.000 \\
MA-VAE & 0.033 & 0.947 & 16.690 & 0.250 \\
\ours & 0.015 & 0.953 & 14.938 & 0.101 \\
\bottomrule
\end{tabular}%
}
\end{table}

\begin{figure}[t!]
    \centering
    \includegraphics[width=1.0\linewidth]{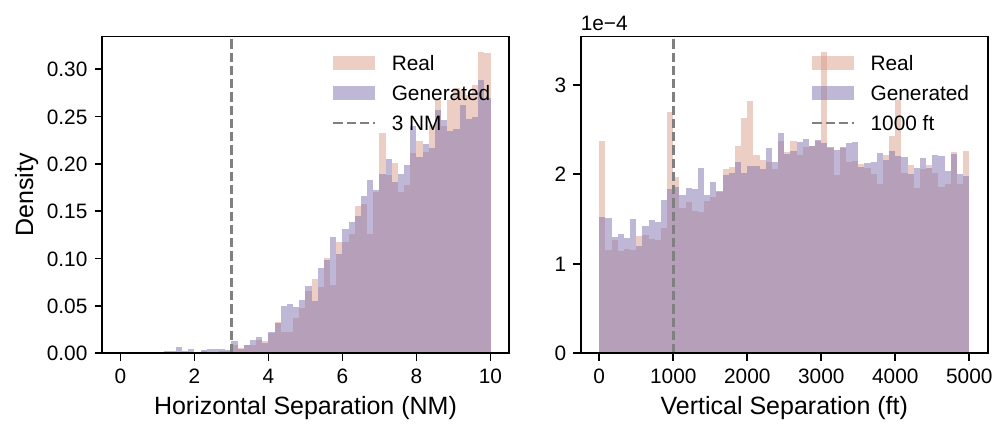}
    \vspace{-18pt}
    \caption{\textbf{Separation patterns in real and generated traffic.}
    Distribution of the minimum horizontal (left) and vertical (right) separation between aircraft pairs. Dashed lines indicate the separation minima.}
    \label{fig:separation}
    \vspace{-5pt}
\end{figure}

\begin{figure}[t!]
    \centering
    \includegraphics[width=1.0\linewidth]{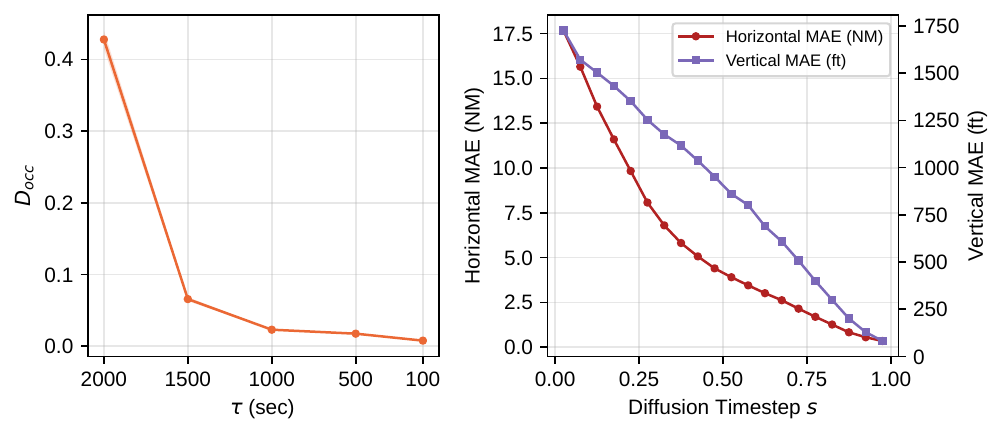}
    \vspace{-18pt}
    \caption{\textbf{Traffic fidelity across arrival stages and diffusion timesteps.}
    JSD between real and generated traffic within each RTA ($\tau$) range (left). Horizontal and vertical mean absolute error between \ours's predicted and true positions as a function of the diffusion timestep $s$ (right).}
    \label{fig:tau_t_curve}
    \vspace{-8pt}
\end{figure}

\begin{figure*}[t!]
    \centering
    \includegraphics[
    width=\textwidth,
    trim={0 2pt 0 2pt},
    clip
]{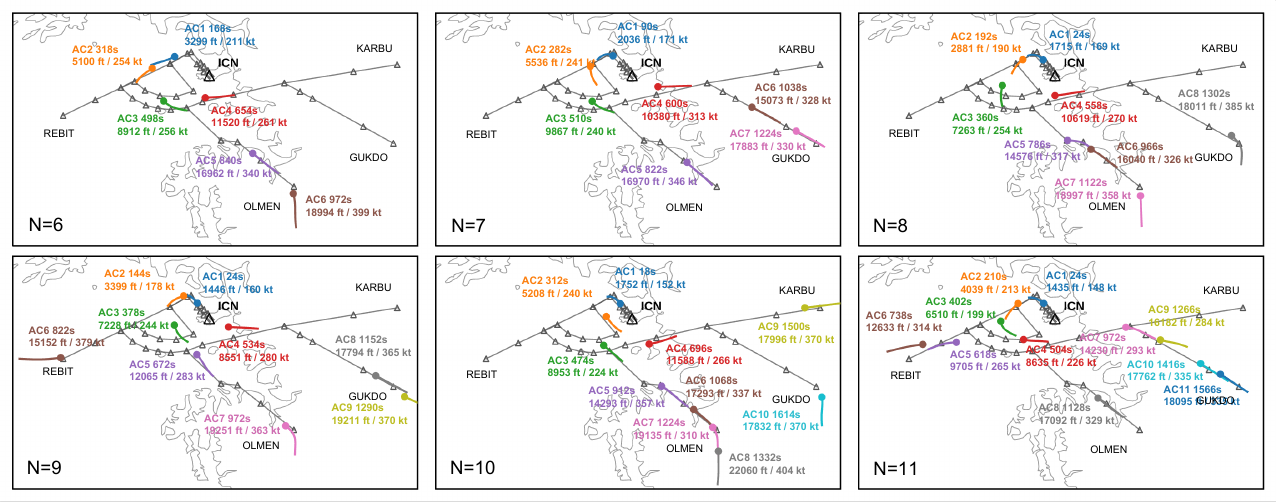}
    \vspace{-18pt}
    \caption{\textbf{Generated traffic scenes with varying numbers of aircraft.}
    Examples of scenes generated by \ours for $N=6$ to $N=11$. Each panel shows one scene over the ICN terminal area, with STARs and runways shown in gray for reference. Aircraft are labeled AC1 to AC$N$ in landing order, and each label reports the RTA $\tau$, current altitude, and ground speed.}
    \vspace{-5pt}
    \label{gen_scene}
\end{figure*}

\begin{figure*}[t!]
    \centering
    \includegraphics[
    width=\textwidth,
    trim={0 2pt 0 2pt},
    clip
]{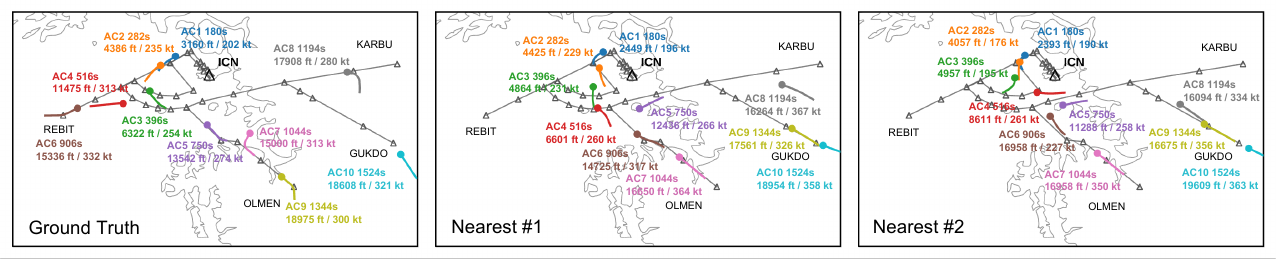}
    \vspace{-18pt}
    \caption{\textbf{Generated alternatives under the same RTA condition.}
    Comparison between a recorded traffic scene and the closest scenes generated by \ours under the same $\boldsymbol{\tau}$.}
    \vspace{-5pt}
    \label{gen_scene2}
\end{figure*}

\ours reduces $D_{\mathrm{occ}}$ by more than half (0.015 vs.\ 0.033) and lowers SVR from 0.250\% to 0.101\% compared with MA-VAE, showing closer agreement with the spatial distribution and inter-aircraft separation observed in real traffic. Both models achieve $R_{\mathrm{nov}}$ close to one (0.953 and 0.947), indicating that their generated scenes remain distinct from the training data. MA-VAE shows greater diversity under a fixed $\boldsymbol{\tau}$ (16.690 NM vs.\ 14.938 NM), but also a substantially higher $D_{\mathrm{occ}}$, suggesting that some of this variation extends beyond the spatial patterns observed in real traffic. Because diversity has no direct reference value, we interpret it mainly as evidence that \ours does not collapse to a single scene for a given condition rather than as a metric for ranking the two models.

The performance difference may be explained by how the two models generate air traffic scenes. During training, MA-VAE obtains agent latents from an encoder that observes the entire scene, so their joint configuration reflects the interactions among aircraft. At generation time, however, each agent latent is sampled independently from $\mathcal{N}(\mathbf{0},\mathbf{I})$. Even if each individual latent follows the prior, independent sampling does not preserve the joint arrangement present during training, leaving the decoder to reconstruct unrealistic inter-aircraft relationships from latent combinations that have not been seen in historical data. In contrast, \ours starts from the same Gaussian prior it is trained to transport and follows the learned joint velocity field over multiple aircraft at inference, reducing the mismatch between training and generation. This may contribute to its closer agreement with real traffic.

Figure~\ref{fig:separation} compares the distributions of minimum horizontal and vertical separation for every aircraft pair in real and generated traffic scenes. The distributions closely match, showing close agreement in both separation violations and overall inter-aircraft spacing. In the real data, pairs falling below the minimum are less frequent horizontally than vertically, indicating that horizontal separation is more consistently maintained under real-world ATC operations, a pattern also reproduced by \ours. Matching these separation patterns is important because inter-aircraft spacing affects both safety and the level of traffic an airspace can accommodate.

We report two additional analyses in Figure~\ref{fig:tau_t_curve}. The left panel groups aircraft by RTAs and computes $D_{\mathrm{occ}}$ within each group, showing how closely generated traffic matches real traffic at different stages of arrival. For RTAs below about 1000 s, $D_{\mathrm{occ}}$ remains near 0.02 with little variation as trajectories converge toward common arrival paths near the runway. $D_{\mathrm{occ}}$ increases mainly for aircraft farther from the runway, where later RTAs allow a wider range of plausible trajectories. The right panel shows the horizontal and vertical mean absolute error between predicted and true scenes as a function of the diffusion timestep $s$. Both decrease monotonically from $s=0$ to $s=1$, indicating that the model progressively refines the scene from noise throughout denoising.

\subsection{\textbf{Qualitative Results}}
Figure~\ref{gen_scene} shows traffic scenes generated by \ours for varying numbers of simultaneous aircraft ($N=6$ to $N=11$). For each aircraft, we report the conditioning RTA $\tau_i$, together with the current altitude and ground speed. Across all panels, \ours distributes aircraft over the four arrival fixes (REBIT, OLMEN, GUKDO, and KARBU), although the conditioning does not specify which fix to use. The scenes also show coherent arrival sequencing: aircraft with earlier RTAs are closer to the runway, lower, and slower. Even when deviating from the standard terminal arrival routes (STARs), aircraft form common arrival flows resembling traffic patterns shaped by ATC vectoring.

\begin{figure*}[t!]
    \centering
    \includegraphics[
    width=\textwidth,
    trim={0 2pt 0 2pt},
    clip
]{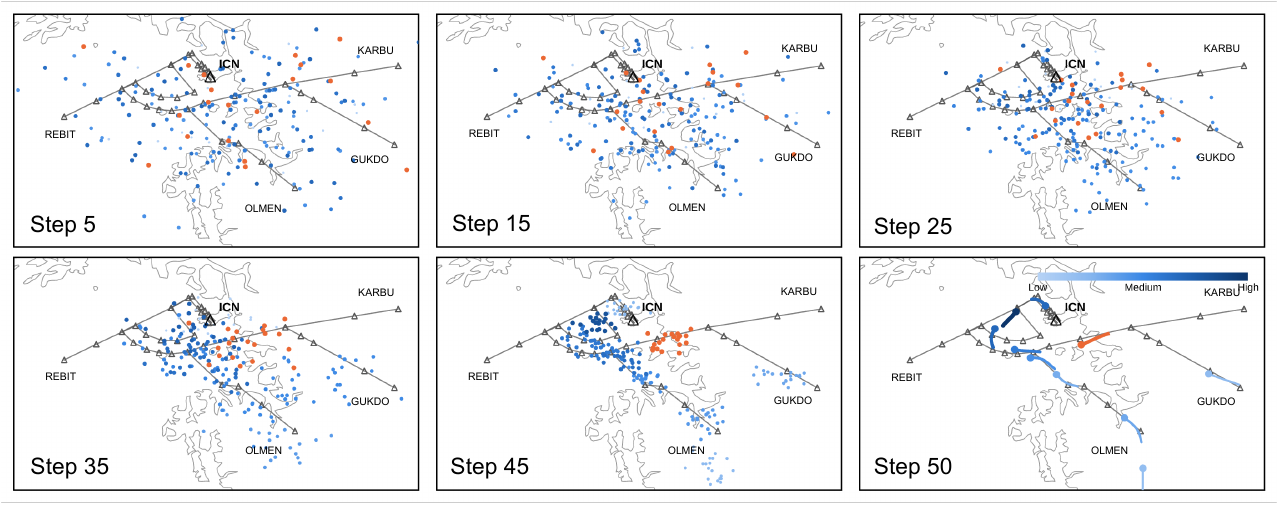}
    \vspace{-18pt}
    \caption{\textbf{Agent attention throughout the denoising process.} The query aircraft is shown in orange, while the other aircraft are colored according to the attention they receive from the query. Intermediate states are shown as points rather than trajectories because they remain noisy during early denoising steps.}
    \label{attn}
\end{figure*}

Particularly, in the generated scene with $N=9$, AC5, AC6, and AC7 have their RTAs separated by approximately 150 seconds. Despite this similar temporal spacing, the spatial distance between AC5 and AC6 is much smaller than that between AC6 and AC7, suggesting that AC6 takes a longer route to the runway to maintain the landing interval. Such behavior is consistent with traffic patterns observed for REBIT arrivals at ICN. Because REBIT is located closer to the airport than the other arrival fixes, ATCs often vector aircraft entering through REBIT to maintain separation and merge them into the arrival stream, increasing their path length and time in the terminal airspace. This example suggests that \ours captures not only the spatial distribution of traffic but also operational patterns induced by ATC vectoring.

To further assess whether \ours generates alternative traffic realizations under the same condition, we take the RTAs $\boldsymbol{\tau}$ from a real scene, generate multiple scenes using that same $\boldsymbol{\tau}$. A useful generative model should represent the range of traffic scenes that could plausibly occur under a given condition, not the one scene that happened to be recorded. Figure~\ref{gen_scene2} shows the ground-truth recorded scene alongside the two generated samples closest to it. Even these closest samples remain visibly different. AC7 provides a clear example: although its RTA is fixed to the same value in all scenes, its trajectory varies substantially. In the ground truth, AC7 turns right and remains outside the main arrival stream while REBIT traffic occupies the airspace near the airport, landing only after that stream has passed. In nearest \#1, by contrast, AC7 joins the existing flow without the same detour. Thus, the same RTAs and arrival sequence can correspond to different plausible trajectory configurations, and \ours generates such alternatives rather than reproducing the recorded scene.

Lastly, Figure~\ref{attn} shows the agent attention scores throughout the denoising process. The query aircraft is shown in orange, while the others are colored according to the attention they receive from the query. Up to about step 25, the scores remain nearly uniform because the scene is still dominated by noise and little relational structure has emerged. As denoising progresses, the attention becomes more differentiated: the query assigns higher attention to aircraft ahead in the arrival sequence and lower attention to those behind. This behavior is consistent with arrival flight operations, where preceding aircraft often have a stronger influence on following traffic. \ours learns this interaction pattern directly during generation without explicit supervision on which aircraft should be attended to.

\section{Conclusion}\label{conclusion}
We present \ours, a context-aware generative framework for multi-agent air traffic scenes. \ours employs MAIFormer layers as the denoising backbone of a diffusion transformer trained with a flow matching objective, enabling the joint generation of interacting aircraft while supporting a variable number of aircraft and conditioning scene generation on their RTAs. Experiments on real-world terminal airspace data show that \ours more closely matches the spatial distribution and inter-aircraft separation of recorded traffic than an MA-VAE baseline, while generating diverse and novel traffic realizations rather than reproducing the training data.

Toward practical deployment, \ours can support experimental studies of how ATCs manage different traffic configurations. By conditioning on the number of aircraft and their RTAs, users can specify an arrival sequence and its temporal spacing and generate multiple traffic scenes that realize it. This enables controlled scenario generation for operational analysis and controller training. \ours can also be used to generate synthetic traffic data for dataset augmentation, particularly for rare or underrepresented arrival scenarios, which may help train more robust and accurate models.

In the very near future, we plan to extend \ours from short-horizon scene generation to full-trajectory generation, toward end-to-end air traffic simulation over longer horizons. We also plan to evaluate physical feasibility more comprehensively with respect to aircraft dynamics and operational constraints. In particular, incorporating explicit separation constraints through constrained sampling is an important direction for future work. These extensions would broaden the applicability of \ours to more realistic air traffic simulation and operational studies.



\bibliographystyle{ieeetr}
{\fontsize{8.59pt}{9.75pt}\selectfont
\bibliography{ref}
}

\end{document}